\documentclass[iop,apjl]{emulateapj}
 \usepackage{amsmath}

\shorttitle{An improved method to test  the DD relation}
\shortauthors{Yang et al.}

\begin{document}
\title{An improved method to test  the Distance--Duality relation}

\author{Xi Yang\altaffilmark{1}, Hao-Ran Yu\altaffilmark{1}, Zhi-Song Zhang\altaffilmark{2}, and  Tong-Jie Zhang\altaffilmark{1,3}}

\email{tjzhang@bnu.edu.cn}

\altaffiltext{1}{Department of Astronomy, Beijing Normal University, Beijing  100875, China}
\altaffiltext{2}{Department of Aerospace Engineering, School of Astronautics, Harbin Institute of Technology (HIT), Harbin Heilongjiang, 150001, China}
\altaffiltext{3}{Center for High Energy Physics,  Peking University, Beijing, 100871, China}

\begin{abstract}
Many researchers have performed cosmological-model-independent tests for the distance¨Cduality (DD) relation.
Theoretical work has been conducted based on the results of these tests. However, we find that almost all of these
tests were perhaps not cosmological-model-independent after all, because the distance moduli taken from a given
type Ia supernovae (SNe Ia) compilation are dependent on a given cosmological model and Hubble constant. In this
Letter, we overcome these defects and by creating a new cosmological-model-independent test for the DD relation.
We use the original data from the Union2 SNe Ia compilation and the angular diameter distances from two galaxy
cluster samples compiled by De Filippis et al. and Bonamente et al. to test the DD relation. Our results suggest that
the DD relation is compatible with observations, and the spherical model is slightly better than the elliptical model
at describing the intrinsic shape of galaxy clusters if the DD relation is valid. However, these results are different
from those of previous work.
\end{abstract}



\section{Introduction}
 The luminosity distance, $D_L$, and the angular diameter distance, $D_A$,  are  both fundamental  observations in astronomy.
They satisfy an important relationship named
the distance-duality (DD) relation \citep{ell2007},  which can be expressed as
\begin{equation}  \label{1}
\frac{D_L}{D_A}(1+z)^{-2}  =  1,
\end{equation}
where  $z$ is the cosmological redshift.
This equation is always valid if and only if  the following three conditions are satisfied \citep{ell1971}:
\begin{itemize}
    \item[$\mathbf{(1)}$] Cosmological models are based on Riemannian geometry;
    \item[$\mathbf{(2)}$] Photons travel along null geodesic;
    \item[$\mathbf{(3)}$] Photon number  is conserved.
  \end{itemize}
The DD relation is violated when all of the above conditions are not satisfied.
 Some non-metric theories  do not meet  the condition $\mathbf{(1)}$ or condition $\mathbf{(2)}$.
Secondly, the DD relation will also be violated  if the photon number is non-conserved \citep{bas2004}.
The sensitivity of detector, dust and exotic theory (e.g. photon decay) can cause non-conservation of photon number.
Therefore, it is necessary to  test whether this relation is valid in our real universe.


The parameter $\eta(z)$  was introduced by previous authors to
test the DD relation \citep{hol2010}, i.e.,
\begin{equation}   \label{2}
\frac{D_L}{D_A}(1+z)^{-2}=\eta(z),
\end{equation}
 where the DD relation holds when $\eta(z)=1$. The way to accomplish a cosmological-model-independent test for the DD relation is to use  the values of  $D_L$ and $D_A$ both from observations \citep{hol2010}. It is important to noted  that the ways to measure $D_L$ and $D_A$ should be cosmology independent and should also be independent of each other.

  Generally,  the data of $D_A$ are got from galaxy cluster samples. Based on observations of  Sunyaev-zeldovich effect (SZE) and X-ray surface brightness from galaxy clusters,  the intrinsic sizes of galaxy clusters can be measured,  which can derive the angular diameter distances of galaxy cluster, $D^{\rm cluster} _A$  \citep{ree2002}.  Moreover, the data of $D_L$ can be obtained from SNe Ia sample compilations.  Plugging these data into Equation (\ref{2}), the DD relation will be test \citep{deb2006}.

    However, Using this method to test the DD relation is inappropriate.  \cite{uza2004} pointed out that the the SZE effects and X-ray techniques are related with the DD relation, which means that the observations of angular diameter distance, $D_A^{\rm cluster}$, and the true angular  diameter distance, $D_A$,   have the follow relation,
    \begin{equation}
     D^{\rm cluster} _A= \eta^2 D_A. \label{cluster}
    \end{equation}
  The above equation (actually, their equation is  slightly  different from ours because of different definition of $\eta$) means $D^{\rm cluster} _A= D_A$ only when the DD relation holds ($\eta=1$). Therefore, $D_{A}^{\rm cluster}$ should not be  put directly into Equation (\ref{2}) to test the DD relation.  \cite{hol2010} plugged Equation (\ref{cluster}) into Equation (\ref{2}) to get $ D^{\rm cluster}_{A}(1+z)^2/D_{L}= \eta(z)$. And then,  they put SNe Ia data $D_{L}$ and $D^{\rm cluster}_{A}$ from two galaxy cluster samples compiled by  \cite{def2005}  and  \cite{bon2006} show that constrain $\eta(z)$,  with assuming  $ \eta(z)=1+\eta_1 z$ and $\eta(z)=1+\eta_2 z/(1+z)$.  Their results show that the DD relation can be accommodated at $2\sigma$ CL for elliptical model and  cannot be accommodated for spherical model even at $3\sigma$ CL.  In subsequent works, \cite{li2011} and \cite{men2012} obtained the conclusions that DD relation is accommodated  at the $1\sigma$ CL  for the elliptical model, and it cannot be accommodated event at the $2\sigma$ for the spherical model. So these results concluded that the elliptical model is better than spherical model. \cite{lim2011} suggested that the deviation of $\eta$ from $1$ may indicate that some breaks on fundamental physical theories.

   However, we suggest that  the  SNe Ia data cannot also  be put directly into Equation (\ref{2}) to constrain $\eta(z)$,
   because the distance modulus, $\mu$, of SNe Ia data depend on cosmological model and the selection of Hubble constant, $H_0$. Therefore, their works may need to be improved.

    In this letter, we make a little change on  the distance estimate procedure of SALT2 \citep{guy2007}, and then, we perform an improved cosmological-model-independent test for the DD relation.


 This letter is organized as follows: Section \ref{howtomu}  briefly introduces the approach for getting distance modulus of SNe Ia.  In Section \ref{sam},
we briefly describe the SNe Ia data \citep{ama2010}  and the angular diameter distances data \citep{def2005,bon2006}.
 In  Section \ref{new}, we propose a new cosmological-model-independent method to test the DD relation and get  results.  Finally, the discussions and conclusions are given in Section \ref{dis}.



\section{A brief introduction of the approach for  getting distance modulus of SNe Ia}  \label{howtomu}

 In astronomy, astronomers use SNe Ia as  a secondary standard candle to measure luminosity distance, because the
peak luminosity of light curve (the graph of luminosity as a function of time) of all  SNe Ia  are nearly  identical. In other words, their peak absolute magnitude $M_{\rm max}$ are nearly identical. Assuming a Cepheid variable and a SNe Ia share  a same host galaxy, one can use  the Cepheid variable to measure the luminosity distance $D_L$ of the host galaxy. And then, combing  the peak magnitude $m_{\rm max}$ of the SNe Ia with  the formula of  distance modulus
\begin{equation}
u=5\lg{D_L}-5=m_{\rm max}-M_{\rm max},  \label{u}
 \end{equation}
the peak absolute magnitude $M_{\rm max}$ of arbitrary SNe Ia can be easily  obtained, because every SNe Ia have an almost same  $M_{\rm max}$. Therefore the luminosity distance of arbitrary SNe Ia can be obtained  if its $m_{\rm max}$ is known. However, the peak luminosity of SNe Ia is not exactly same, which is related to the shapes and
colors of the light curves of SNe Ia \citep{guy2005}, and the  extinction effects the magnitude $m_{\rm max}$.
So Equation (\ref{u}) need to be modified.  Many fitters (SALT \citep{guy2005}, SALT2 \citep{guy2007}, MLCSC2K2 \citep{jha2007}) have been proposed to parameterize the light curves of SNe Ia and  its distance modulus can be obtained.

We take the light curves fitter SALT2 as an example to illustrate the process of obtaining distance modulus \citep{guy2007}. \cite{guy2007} modified Equation (\ref{u}) by adding perturbations of shapes and colors to get
\begin{equation}
 \mu_{B}(\alpha, \beta,M_B)=m^{\rm max}_{B}-M_{B}+\alpha x-\beta c, \label{umodify}
\end{equation}
where $m^{\rm max}_{B}$ is the rest-frame peak magintude of B bands, $x$ is stretch factor, which describes the effects of shapes of light curves on $\mu$, and $c$  is color parameter,  which representations the influences of the intrinsic color   and reddening by dust on $\mu$. These three parameters  can be obtained by  fitting the light curves of SNe Ia. Thus, they are independent of cosmological model.  Absolute magnitude $M_{B}$, $\alpha$, and $\beta$ are nuisance parameters, which will be fitted by minimizing the residuals in Hubble diagram that given by a cosmological model. For example,  \cite{ama2010} used  method above and $\chi^2$ minimization to constrain $\Omega_{M}$, $\omega$ and    to get the Union2  compilation. The formula of $\chi^2$ minimization is
\begin{align}
&\chi^2(\alpha,\beta,M_B) \nonumber \\
&=\sum_{\rm SNe}\Big[\frac{\mu_{B}(\alpha,\beta, M_B;z)-\mu^{\rm theory}(\Omega_{M},\Omega_{\omega},\omega;z)}{\sigma_{\rm total}}\Big]^2, \label{uchi2}
\end{align}
where $\mu^{\rm theory}(\Omega_{M},\Omega_{\omega},\omega;z)$ is obtained from the  $\omega$CDM model. The best-fitted values of $\alpha$, $\beta$ and $M_B$ are got by minimizing  $\chi^2$, and then, $\mu$ are obtained.  Obviously, $\mu$ is strongly dependent on $\omega$CDM model  because its values is got from Equation (\ref{uchi2}). One point should be noticed that $\mu^{\rm theory}(\Omega_{M},\Omega_{\omega},\omega;z)$  contains a constant term $5\lg H_0$. For example,  the formula of $D_L$  for flat $\Lambda$CDM model reads
\begin{equation}
D_{L}=\frac{c(1+z)}{H_0}\int^{z}_{0}\frac{dz'}{[\Omega_{M}(1+z')^3+ (1-\Omega_{M})]^{1/2}}, \label{Ho}
\end{equation}
 where $c$ is speed of light and $H_0$ is Hubble constant. One can easily see the constant term $5\lg H_0$ by putting this equation into Equation \ref{u}.

Therefore, $M_B$ is degenerate with $H_0$ because they are constants and have a same status in Equation (\ref{uchi2}).   Using the method of minimizing $\chi^2$,  \cite{ama2010} just got the best fitted values of $M_B-5\lg H_0$, $\alpha$, and $\beta$. and  they chose $H_0 = 70 \rm km \ s^{-1} \ Mpc^{-1}$  to get $M_B$ and $\mu$. It is obviously that $\mu$ of Union2 SNe Ia compilation is dependent on the choice of  $H_0$ and $\omega$CDM model. The distance modulus $\mu$ of other SNe Ia samples (e.g. Constitute \citep{hic2009}, Davis07 \citep{dav2007})  are obtained in the similar way. So  $\mu$ of  SNe Ia   samples   depends on cosmological model and  $H_0$.  It is important to note that the arbitrary selection of $H_{0}$ does not effect the restriction on $\Omega_M$ and $\omega$, because they   used
$m_{B}^{\rm max}$, $c$, and $x$ rather than $\mu$ to constrain $\Omega_M$ and $\omega$.

Thus, it is inappropriate to directly use distance modulus $\mu$ of SNe Ia sample to  test the DD relation. In this letter, we bypass $\mu$ and directly use $m^{\rm max}_{B}$, $x$, and $c$ of Union2 sample to test the DD relation.  Marginalizing $M_{B}$, $\alpha$, and $\beta$, the probability distribution of $\eta$ is got.

\section{Samples} \label{sam}
In order to test the DD relation, we need $D_L$  and $D^{\rm cluster}_A$ data both from cosmological-model-independent
measurement. For $D_L$, we use  Union2 SNe Ia data \citep{ama2010}, which contains 557 well-measured SNe Ia. For $D^{\rm cluster}_A$, we employ SZE and X-ray observations of two galaxy cluster samples: elliptical model sample \citep{def2005} and spherical model sample \citep{bon2006}.    Elliptical model sample was compiled by  \cite{def2005} with an isothermal elliptical $\beta$ model, which contains 18 galaxy clusters compiled by \cite{ree2002} and 7 galaxy clusters compiled by  \cite{mas2001}. Assuming  that the distribution of cluster plasma and dark matter is  hydrostatic equilibrium  and spherical geometry, \cite{bon2006} (see its Table 2) complied the spherical model sample that  includes 38 galaxy clusters. In principle,  giving  a  $D^{\rm cluster}_A$, one should select a $D_L$ of  SNe Ia data point that  shares  the same redshift $z$ with the given data point of $D^{\rm cluster}_A$  to get  $\eta$. However,  the above condition usually cannot be satisfied in reality. So we use the selection criteria, as in  \cite{hol2010}, $\Delta z= |z_{\rm SNe}-z_{\rm cluster}|<0.005$ to select SNe Ia point. If  there are at least two SNe Ia data satisfy this criteria for a given $D^{\rm cluster}_A$,  we select the SNe Ia data whose $\Delta z$ is smallest. The selection criteria can be satisfied for all galaxy clusters data except the cluster CL J1226.9 +3332 from spherical model sample \citep{bon2006}, which  just gives the $\Delta z= 0.005$. We keep this cluster data point in our analysis.
\section{New  test for the DD relation and results} \label{new}
Combing Equation (\ref{2}) with Equation (\ref{cluster}), we get
\begin{equation}
D_{L}=\eta(z)^{-1}D^{\rm cluster}_{A} (1+z)^2, \label{end}
\end{equation}
and then, we define
\begin{equation}
  \mu_{\rm cluster }(\eta;z) = 5\lg{\big[ \eta(z)^{-1}D^{\rm cluster}_{A} (1+z)^2\big]}-5, \label{defineu}
\end{equation}
which is  the distance modulus of a   galaxy cluster data point. Because $D_{A}\simeq D_{L}$ when $z\rightarrow0$, we parameterize $\eta(z)$ in the following form as \cite{hol2010},
\begin{equation}
\eta(z)=1+\eta_{0}z,
\end{equation}
where $\eta_0$ is a constant. We only use this functional form in our analysis because other forms of $\eta(z)$ can approximate to this form by Taylor expansion when $z<1$ (the redshift of all data points in our analysis  is smaller than $1$).Now, we use $\chi^2$ minimization to constrain $\eta_0$,
\begin{align}
&\chi^2(\alpha, \beta, M_B, \eta_0) \nonumber \\
& =\sum_{i}\frac{\big[\mu_{B}(\alpha,\beta,M_B;z_i)-\mu_{\rm cluster}(\eta_0;z_i)\big]^2}
{\sigma^2_{\rm total}(z_i)}, \label{newchi2}
\end{align}
where $\mu_{B}(\alpha,\beta,M_B;z)$ of SNe Ia  comes from Equation (\ref{umodify}), $\mu_{\rm cluster}(\eta_0;z)$ of galaxy cluster is given by  Equation  (\ref{defineu}), and the uncertainty $\sigma^2_{\rm total}(z)$ is given by
\begin{equation}
 \sigma^2_{\rm total}(z)=\sigma^2_{m}(z)+\alpha^2\sigma^2_{x}(z)+\beta^2\sigma^2_{c}(z)
+\Big[\frac{5}{\ln10}\cdot\frac{\delta_{D_A}(z)}{D^{\rm cluster}_A}\Big]^2,
\end{equation}
where  $\sigma_{m}$,  $\sigma_{x}$, $\sigma_{c}$, and $\delta_{D_A}$ are the errors of $m^{\rm max}_{B}$, $x$, $c$, and $D^{\rm cluster}_A$ respectively.

Inserting data points ($m^{\rm max}_{B}$, $x$, $c$, $\sigma_{m}$, $\sigma_{x}$, $\sigma_{c}$) of SNe Ia Union2  and data points ($D^{\rm cluster}_A$, $\delta_{D_A}$) of galaxy cluster samples into Equation (\ref{newchi2}),  $\chi^2(\alpha, \beta, M_B, \eta_0)$ is got. And then, the joint probability density of these parameters  can  be get, $P(\alpha, \beta, M_B, \eta_0)=A\exp(-\chi^2/2)$, where $A$ is a normalized coefficient, which
makes $\iiiint{ P \  d\alpha d\beta dM_B d\eta_0} =1$. By Integrating over $\alpha$, $\beta$, and $M_B$, the probability
distribution function of $\eta_0$ is gained, i.e., $P(\eta_0)= \iiint^{+\infty}_{-\infty}{P(\alpha, \beta, M_B, \eta_0) \  d\alpha d\beta dM_B}$.

  We adopt iterative  method to calculate $P(\eta_0)$ with step size 0.01 for all parameters. In principle, we should calculate all the values of $\chi^2$   of $\alpha$, $\beta$,  $M_B$,  and $\eta_0$ from $-\infty$ to $+\infty$. Obviously, It is impossible to do that. We just calculate   the values of  $\chi^2(\alpha, \beta, M_B, \eta_0)$  for these parameters in $3\sigma$ interval instead of infinite interval. And then,  we get $\chi^2$ , $P(\alpha, \beta, M_B, \eta_0)$, and $P(\eta_0)$, with  $P(\eta_0)\propto \sum_{i} \sum_{j} \sum_{k}{P\big(\alpha(i), \beta(j), M_B(k), \eta_0\big)}$, where $i$, $j$, $k$ run over all the data points for  $\alpha$, $\beta$, and $M_B$ in $3 \sigma $ interval with step size $0.01$ respectively.

   The  next step is to use the equation, $\Delta \chi^2= \chi^2-\chi^2_{\rm min}$, to constrain the one dimensional CL of parameter with $1$ and $4$ level \citep{pres1992}, where $\chi^2_{\rm  min}$ is the minimum of $\chi^2$.
     For example, if we want to calculate  $1 \sigma $ and $2 \sigma$ CL of $\eta_0$. We just  need to find out the data points of $\eta_0$ which satisfy $\Delta \chi^2 \leq1$  and $ \Delta \chi^2 \leq 4$ respectively.
\begin{figure*}
\begin{center}
$\begin{array}{l}
\includegraphics[width=1\textwidth,height=0.32\textwidth]{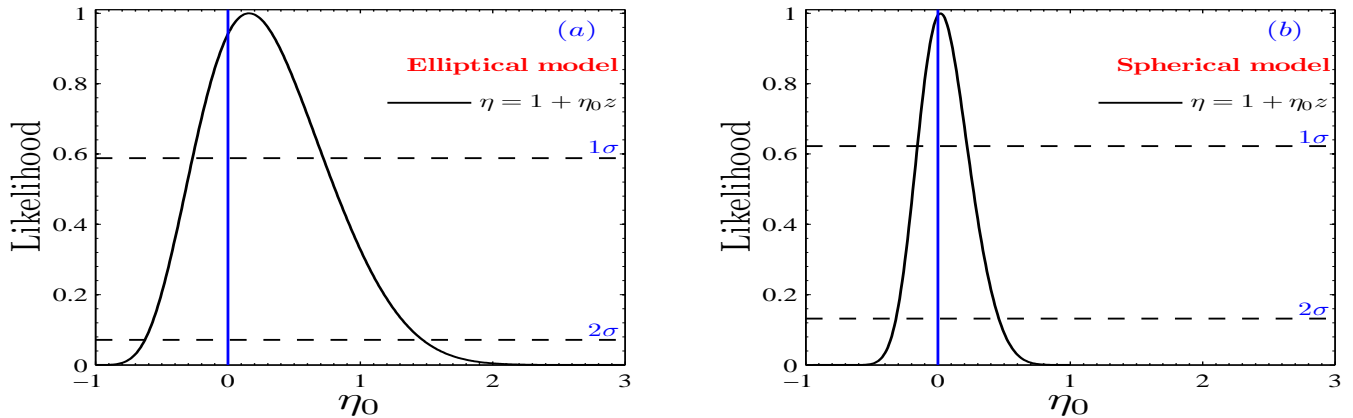}
\end{array}$
\end{center}
\caption{
$(a)$ Likelihood distribution  function (LDF) of $\eta_0$ for elliptical model.
$(b)$ LDF of $\eta_0$ for spherical model.}
\label{f1}
\end{figure*}
\begin{figure*}
\begin{center}
$\begin{array}{l}
\includegraphics[width=1\textwidth,height=0.32\textwidth]{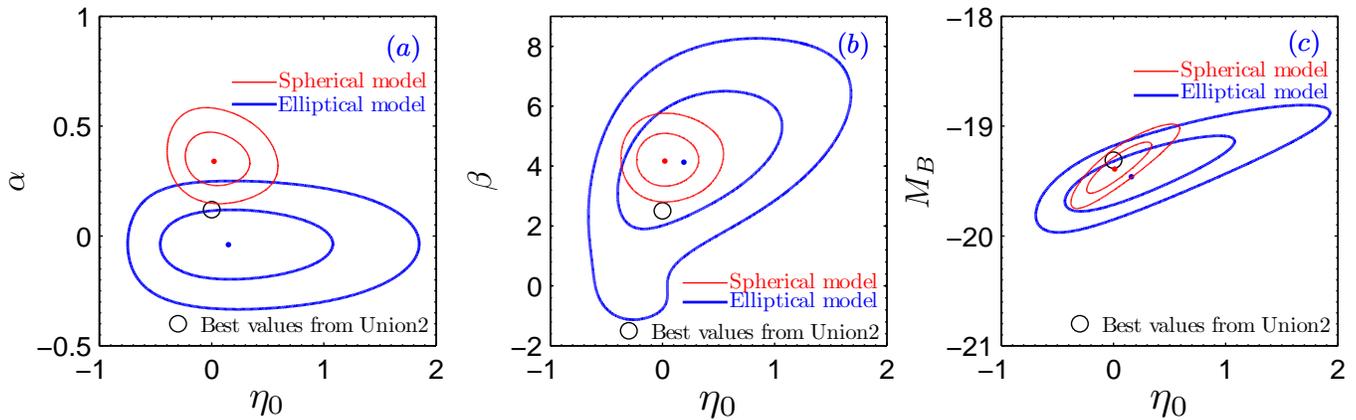}
\end{array}$
\end{center}
\caption{ Confidence regions at $1 \sigma$ and $2 \sigma$ level  on planes   $(\eta_0, \alpha)$, $(\eta_0, \beta)$, and $(\eta_0,M_B)$ respectively. The red  and blue contours are derived from the spherical model and the elliptical model sample respectively.   $``o"$ represents the best values in Union2 by assuming $\eta_0=0$.
}
\label{f2}
\end{figure*}

    Using the above procedure, Combing Union2 SNe Ia sample \citep{ama2010} and  elliptical model sample \citep{def2005} to calculate $\chi^2$, we get that the best-fitting values of $M_{B}$, $\alpha$, $\beta$, and $\eta_0$ are $-19.45$, $-0.03$, $4.05$, and $0.18$ respectively. And then, we marginalize $\alpha$, $\beta$, and $M_B$ by integrating over them to get the likelihood  distribution of $\eta_0$ and obtain that $\eta_0 = 0.16_{-0.39}^{+0.56}$ at $1 \sigma$ CL and $\eta_0 = 0.16_{-0.70}^{+1.31}$ at $2 \sigma$ CL. The likelihood distribution of $\eta_0$ is shown in Figure \ref{f1}(a), For spherical model sample \citep{bon2006},   in the same way,  we get that the best-fitting values of $M_{B}$, $\alpha$, $\beta$, $\eta_0$ are $-19.38$, $0.34$, $4.13$, and $0.02$  respectively, and we obtain that $\eta_0 =0.02_{-0.17}^{+0.20}$ at $1 \sigma$ CL and $\eta_0= 0.02_{-0.33}^{+0.44}$ at $2 \sigma$ CL. The probability distribution of $\eta_0$  is shown  in Figure \ref{f1}(b). Figure \ref{1} shows that the DD relation $(\eta_0 =0)$  can be better satisfied at $1 \sigma$ CL for spherical model and elliptical model.

       The reason why we get a different result from previous works \citep{hol2010, li2011,men2012} is that we perform a  different procedure. In the previous method, $D_{L}$ is directly obtained from distance modulus $\mu$ of SNe Ia.  Instead, in our analysis, $D_{L}$ is  not provided by distance modulus $\mu$ but by original data $m_{B}, x, c$.

       For example, in the work of \cite{li2011} ,  data $D_L$ are taken from the Union2 SNe Ia compilation. In Union2 \citep{ama2010}, $\mu$ are dependent on the best fitting values of  $M_B$, $H_0$, $\alpha$, and $\beta$. Obviously, the values of  $\alpha$ and $\beta$ are derived from a fit to $\omega$CDM model, and  $M_B$ is determinated by choice $H_0 = 70 \rm km \ s^{-1} \ Mpc^{-1}$. The best fitting values of $M_B$, $\alpha$, and $\beta$ in Union2 are $-19.31$, $0.12$, and $2.51$ respectively. In our analysis, the best fitting values of  $M_B$, $\alpha$, and $\beta$
       are $-19.45$, $-0.03$, $4.05$ and $-19.38$, $0.34$, $4.13$  for elliptical model and spherical model respectively.
       Because the difference of  best fitting values of parameters  shift the peak of probability distribution of $\eta_0$, our result will be different from \cite{li2011} inevitably. Their results show that the DD
       relation can be  accommodated for elliptical model at $1  \sigma$ ($\eta_0=-0.07_{-0.19}^{+0.19}$) but cannot be
       accommodated for spherical model even at $2\sigma$ ($\eta_0=-0.22_{-0.11}^{+0.11}$). Moreover,  we marginalize parameters $M_B$,  $\alpha$, and $\beta$ by integrating over them to plot $P(\eta_0)$. Obviously, marginalization will broaden the profile of $P(\eta_0)$. Therefore, this operation may make the DD relation hold at $1 \sigma$  by expanding the $1\sigma$ CL of $\eta_0$.  Hence, Because of the difference of best fitting values of parameters and  marginalization,  we get the different conclusion that the DD relation is compatible with observations at $1 \sigma$.

 Furthermore, we plot the two dimensional contour of $\eta_0$ vs $\alpha$ ,  $\eta_0$ vs $\beta$, and  $\eta_0$  vs $M_B$
 for the two galaxy cluster samples (see figure (\ref{f2})) to see whether if there is a degeneracy effects the DD relation. In figure (\ref{f2}), The red  and blue contours are derived from  spherical model and  elliptical model respectively. $``o"$ indicates the best fitting values of $\alpha$, $\beta$, $M_B$ $(-19.31, 0.12, 2.51)$  by assuming $\eta_0=0$ in Union2 sample. The best values of planes ($\eta_0$, $\alpha$), ($\eta_0$, $\beta$), and ($\eta_0$, $M_B$) are $(0.02, 0.34)$, $(0.02, 4.17)$, $(0.01, -19.39)$  for red contour respectively, and  $(0.15, -0.04)$, $( 0.19, 4.13)$,  $(0.16, -19.46)$ for blue contour respectively. There is no distinct degeneracy effects on $\eta_0$ because all the planes have similar intervals for $\eta_0$ at $1 \sigma$ and $2 \sigma$.  With the same operation on $\eta_0$, we obtained that  $1 \sigma$ CL  of  parameters $M_B$, $\alpha$, and $\beta$  respectively are $-19.37^{+0.14}_{-0.16}$, $0.34_{-0.06}^{+0.08}$,  $4.19_{-0.62}^{+0.58}$  for spherical model, and
  $-19.42^{+0.20}_{-0.24}$, $-0.04_{-0.09}^{+0.10}$, $4.35_{-1.73}^{+1.20}$ for elliptical model sample. In Union2 SNe Ia sample \citep{ama2010}, they are  $-19.31^{+0.014}_{-0.014}$, $0.121_{-0.007}^{+0.007}$, $2.51_{-0.007}^{+0.007}$ respectively.  Comparing these three sets of data, one find that just the confidence level of $M_B$ can be compatible with each other.  For $\alpha$,
   these samples  fail to  be compatible with each other at $1\sigma$. For $\beta$,  the two galaxy cluster samples get $\beta\sim4.10$  which is larger than $2.51$ given by  SNe Ia sample Union2., one can see these conclusions from Figure (\ref{f2}). The reasons why these three sets of  data are different from each other may be worth thinking about.

\section{Discussions and conclusions} \label{dis}
\cite {hol2010} proposed a cosmological-model-independent method to test the DD relation. Using this method, many works \citep{ nai2011,nai2012,yan2013} have been done.   However, we indicates that their method may depend on the selection of $H_0$ and cosmological model. Different choices of $H_0$  will lead to different constraints on $\eta_0$, and one cannot
eliminate this effect by marginalizing $H_0$, because $M_B$ is degenerate with $H_0$.

Therefore,  we improve their  method to perform a new test for the  DD relation again.
 In the previous method \citep{hol2010}, $D_{L}$ is directly obtained from distance modulus $\mu$ of SNe Ia.  Instead, in our analysis, $D_{L}$ is  not provided by distance modulus $\mu$ but by original data $m_{B}, x, c$. In this way, we  do not need
 a given cosmological model and any information about $H_0$. Hence, our test is independent of cosmological model and $H_0$.

Our results show that the DD relation can be accommodated at $1\sigma$ CL  well for both elliptical model and spherical model, and   spherical model is slightly better than the elliptical  model if the DD relation is valid. This results, however, are different from  the   previous works  \citep{hol2010, li2011, men2012}. In their works,  the DD relation at most can be accommodate at $1 \sigma$  CL marginally for elliptical model sample, and the DD relation just can  be barely satisfied at $3\sigma$ CL for spherical model sample.  So they concluded that the elliptical model is better than the spherical model in describing the intrinsic shape of galaxy clusters, and some works \citep{lim2011, nai2012} used the  deviation of $\eta(z)$ from $1$ to search for the news physics. However,  from Figure \ref{f1}, one can see clearly that the DD relation is compatible with both elliptical model and spherical model very well at $1 \sigma$ CL, Thus, the  conclusions obtained by the previous works  may need to be treated with caution.

 Furthermore, one  thing need to be noted that the  best fitting values of parameters $M_B$, $\alpha$ and $\beta$  in different samples are some different. Just $M_B$ can be compatible with each other. This at least means the luminosity distance from these samples are similar because stretch factor $x_1$ and color factor $c$ just are perturbation, even $M_B$ in Union2  depends on $H_0$.  For $\alpha$,  elliptical model sample gives the smallest values of $\alpha$, and spherical gives the biggest values of $\alpha$. For $\beta$,  the two galaxy cluster samples get $\beta\sim4.10$  which is larger than $2.51$ given by  SNe Ia sample Union2. This means that the color  factor make a bigger effects on $\mu$ in our analysis than in standard cosmological analysis. Maybe these differences is produced by some unknown physics effects, inaccuracy or
 few number data points of galaxy cluster samples. Nevertheless,   using more galaxy cluster samples or other methods which are independent of cosmological model to constrain $\alpha$, $\beta$ and $M_B$, and researching what cause these differences of them   may be worth doing in the future.

\acknowledgments
We thank the anonymous referees whose suggestions greatly helped us improve this paper.
X. Y. is very grateful to Jian-Chuan Zheng, Ji-Long Chen and Peng-Xu Jiang  for their kind helps.
This work was supported by the National Science Foundation of China (Grants No. 11173006), the Ministry of Science and Technology National Basic Science program (project 973) under grant No. 2012CB821804, and the Fundamental Research Funds for the Central Universities.


\end{document}